\documentclass[mathleft
]{an}
\usepackage{graphicx,amssymb}
\usepackage{times}
\usepackage{natbib}
\newcommand{\ltsima} {$\; \buildrel < \over \sim \;$}  
\newcommand{\gtsima} {$\; \buildrel > \over \sim \;$}  
\newcommand{\lta} {\lower.5ex\hbox{\ltsima}}  
\newcommand{\gta} {\lower.5ex\hbox{\gtsima}}

\newcommand{\ergs}{\>{\rm erg}\,{\rm s}^{-1}}

\newcommand{\loiii}{L$_{\rm{\tiny{ [O~III]}}}$}

\newcommand{\forb}[2]{\mbox{$[{\rm #1\, #2}]$}}
\newcommand{\oiii}{\forb{O}{III}}

\begin{document}

\Pagespan{789}{}
\Yearpublication{2006}%
\Yearsubmission{2005}%
\Month{11}%
\Volume{999}%
\Issue{88}%

\title{The new class of FR~0 radio galaxies}

\author{Ranieri D. Baldi\inst{1,2}\fnmsep\thanks{Corresponding author:
  \email{r.baldi@soton.ac.uk}\newline}
\and  Alessandro Capetti\inst{3} \and Gabriele Giovannini\inst{4,5}
}
\titlerunning{The new class of FR~0 radio galaxies}
\authorrunning{R.~D. Baldi, A. Capetti \& G. Giovannini}
\institute{
Physics Department, The Technion, 32000 Haifa, Israel
\and 
Department of Physics and Astronomy, The University, Southampton SO17 1BJ, UK
\and
INAF - Osservatorio Astrofisico di Torino, Strada Osservatorio 20, I-10025 Pino Torinese, Italy
\and 
Dipartimento di Fisica e Astronomia, Universit\`a di Bologna, via
  Ranzani 1, 40127 Bologna, Italy 
\and INAF-Istituto di Radio Astronomia, via P. Gobetti
  101, I-40129 Bologna, Italy}

\received{30 May 2005}
\accepted{11 Nov 2005}
\publonline{later}

\keywords{Editorial notes -- instruction for authors}

\abstract{Are the FR~I and FR~II radio galaxies representative of
  the radio-loud (RL) AGN population in the local Universe? Recent
  studies on the local low-luminosity radio sources cast lights on an
  emerging population of compact radio galaxies which lack extended
  radio emission. In a pilot JVLA project, we study the
  high-resolution images of a small but representative sample of this
  population. The radio maps reveal compact unresolved or slightly
  resolved radio structures on a scale of 1-3 kpc. We find that these
  RL AGN live in red massive early-type galaxies, with large black
  hole masses ($\gtrsim$10$^{8}$ M$_{\odot}$), and spectroscopically
  classified as Low Excitation Galaxies, all characteristics typical
  of FRI radio galaxies which they also share the same nuclear
  luminosity with. However, they are more core dominated (by a factor
  of $\sim$30) than FRIs and show a clear deficit of extended radio
  emission. We call these sources 'FR0' to emphasize their lack of
  prominent extended radio emission. A posteriori, other compact radio
  sources found in the literature fulfill the requirements for a FR~0
  classification. Hence, the emerging FR0 population appears to be the
  dominant radio class of the local Universe. Considering their
  properties we speculate on their possible origins and the possible
  cosmological scenarios they imply.}

\maketitle

\section{Introduction}

Classical radio catalogs (e.g. 3C, 2Jy...) are set at low radio
frequencies and limited by high flux density. These selection criteria
favor the inclusions of extended steep spectrum sources,
Fanaroff-Riley classes (FR~I and FR~II), where the core emission
contributes $\sim$1\% of their total emission and their sizes are
of tens or hundreds of kpc (e.g \citealt{morganti97}). Furthermore,
they produce a bias, since the high core-dominant sources are
excluded. In fact, when the selection biases used are less severe
(lower flux threshold and/or higher frequency), core dominated
radio-galaxies (RGs), generally too faint to be detected in the
existing low-frequency limited surveys, emerge as the dominant
constituent of the RL AGN population \citep{baldi10a}. Hence, do the
classical FR~I and FR~II RGs represent the real picture of RL AGN?  In
this contribution to the Proceedings, we want to adress this question
in the context of the jet formation in the local Universe.

We investigate the properties of the local RL AGN population, selected
by \citet{best05a} (hereafter B05), a sample of RGs by
cross-correlating the SDSS (DR2), NVSS, and FIRST datasets. This
sample is highly (95\%) complete down to the flux threshold of 5 mJy
and provides a very good representation of RGs in the local Universe,
up to a redshift of $\lesssim$0.3, covering the range 10$^{38-42}$ erg
s$^{-1}$ in radio power. All morphologies are represented, including
twin-jets and core-jet FR Is, narrow and wide angle tails, and FR
IIs. However, most of them $\sim$80 \% are compact, unresolved or
barely resolved at the 5$\arcsec$ FIRST resolution, corresponding to a
limit to their size of $\sim$10 kpc. This sample does not mostly
overlap with previous classical radio catalogs because of their low
radio flux density and is more numerous than 3C sample of a factor of
$\sim$100 in space density.

\begin{figure}
\centering
\includegraphics[scale=0.45,angle=0]{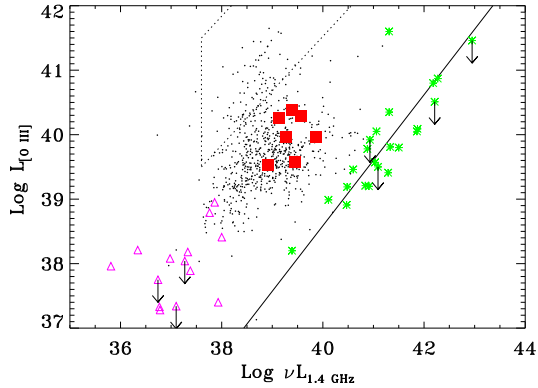}
\caption{{\small{FIRST vs. [O~III] line luminosity (erg
  s$^{-1}$). The small points correspond to the B05 sample. The
  solid line represents the correlation between line and
  radio-luminosity derived for the 3CR/FR~I sample. The dotted lines
  include the region where Seyfert galaxies are found. The empty pink
  triangles are the CoreG, and green stars the 3CR/FR Is. The red
  squares are the FR~0s.}}}
\label{lrlo3} 
\end{figure}

\citet{baldi10a} analyzed their spectro-photometric properties and
they found that they display a strong deficit of radio emission with
respect to their nuclear emission-line luminosity up to a factor 1000,
when compared to FR~I and FR~II matched in line luminosity
(Fig.~\ref{lrlo3}). Most of them live in red massive ($\sim$10$^{11}$
M$_{\odot}$) early-type galaxies (ETGs), with large BH masses
($\gtrsim$10$^{8}$ M$_{\odot}$), and spectroscopically classified as
Low Excitation Galaxies, all characteristics typical of FR~Is.

A similar radio behavior is also noted in the `miniature' RGs,
i.e. Core Galaxies (CoreG), nearby giant ETGs of extremely low radio
luminosity (10$^{20-22}$ W Hz$^{-1}$ at 1.4 GHz)
\citep{balmaverde06core}. Despite their low total radio power, they host
RL nuclei and produces jets on a scale of $\sim$20 kpc, smaller than
the typical size of $\sim$100 kpc of the FR~Is. Their host and nuclear
properties are indistinguishable from the FR~Is (e.g
\citealt{balmaverde06,balmaverde08,baldi09}). However, CoreG show a
core dominance up to $\sim$100 times higher than FR~Is, but similar to
the B05 sample \citep{baldi09}. A high core dominance is
generally interpreted as evidence of Doppler boosting in a radio
source oriented at a small angle with respect to the line of
sight. The correlation found with the core radio power and emission
line luminosity (independent of orientation) indicates that this is a
genuine deficit of extended radio emission and that no geometric
effect is present.

The smaller jets produced by the CoreG compared to FR~Is might be ascribed to
the their lower AGN luminosity, although they have similar
properties. However, the B05 sample, which show a similar radio deficit to the
CoreG, have optical luminosities similar to the FR~Is and their nuclei and
hosts are statistically indistinguishable from those of 3C sources from the
point of view of morphology, color, stellar and BH masses. Therefore, this
radio behavior cannot be ascribed to differences in their hosts, but to other
mechanisms.

In conclusion, the properties of the bulk of the population of RL AGNs
(with a space density $\sim$100 times higher than 3C sources) are
virtually unexplored. This severely limits our ability to understand
their nature and the jet formation in low-luminosity AGN. An
improvement of the radio data is needed to provide a more complete
view of the radio emission phenomenon. For this purpose, we start a
project with the JVLA aimed at study the high-resolution properties of
the RL AGN population, represented by the B05 sample. Here, we
present the results from a pilot sample, largely discussed in
\citet{baldi15a}.

\section{Sample}

We observe with the JVLA at three frequencies (1.4, 4.5, and 7.5 GHz)
twelve objects selected from the B05 sample with the following
criteria: redshift $z<0.1$; the main optical emission lines detected
at least 5 $\sigma$ significance; equivalent width for the
\oiii\ lines larger than 3 \AA.

We will study the RL AGN objects present in this sample, since a
contamination from radio-quiet (RQ) AGN is expected to be present in
the B05 sample. 

\section{Results}
\subsection{Spectro-photometric properties}

We now explore the physical properties of the present sample based on
the optical spectroscopic and photometric information available from
the SDSS survey, similarly to the study performed on the B05 sample by
\citet{baldi10a}.

We estimated the BH masses from the stellar velocity dispersion
adopting the relation of \citet{tremaine02}.  They range from
$\sim$10$^{7}$ to $\sim$10$^{9}$ M$_{\odot}$. The sources are
associated with galaxies with a distribution of masses in the range
10$^{10}$-10$^{11.5}$ M$_{\odot}$.

Studying their concentration index $C_{r}$ (defined as the ratio of
the radii including 90\% and 50\% of the light in the $r$ band) to
carry out a morphological classification of the hosts
(e.g. \citealt{shen03}), either early- and late-type galaxies are
present in the sample. We also use the 4000\AA\ break strength,
$D_n(4000)$ (defined as the ratio between the fluxes in the wavelength
range 3850-3950\AA\ and 4000--4100\AA), sensitive to the presence of a
young stellar population \citep{balogh99}.  A further diagnostic panel
tool enables us to to qualitatively measure the amount of
contamination in radio emission in the galaxy that is due to star
formation. This method is based on the location of a galaxy in the
$D_{n}(4000)$ versus $L_{{\rm 1.4 \,\,GHz}}/M_{*}$ plane, where
$M_{*}$ is the galaxy's stellar mass (Fig. 6 from
\citealt{baldi15a}). Nine sources are above above the curve
corresponding to the prediction of a star formation event lasting 3
Gyr and exponentially decaying. According to B05, their radio emission
is associated with the AGN.

We used the spectroscopic diagnostic diagrams defined by
\citet{buttiglione10} for the 3CR sample to recognize the nature of
their nuclear emission. These diagnostics are formed by pairs of
nuclear emission line ratios to separate active nuclei from
star-forming galaxies (e.g. \citealt{baldwin81}) and, furthermore, to
separate AGN into branches of different excitation level, that is Low-
and High-excitation galaxies (LEG and HEG). So, we classify
the spectra of our sample in nine LEGs and three HEGs (Fig. 1 from
\citealt{baldi15a}).

Based on the spectro-photometric properties, we can distinguish the
presence of two groups. The first group consists of four sources that
are characterized by their low BH masses, mostly $\sim$10$^{7}$
M$_{\odot}$ and their blue color. Their radio and spectrophotometric
properties indicate that they are RQ AGN. The presence of an active
nucleus is evidenced by their optical line ratios and equivalent
widths, which are characteristic of AGN. Nonetheless, three of them
show a substantial contamination from star formation to their radio
emission and are in late-type galaxies.

Conversely, the second group includes eight sources with high BH
masses ($\gtrsim$10$^{8}$ M$_{\odot}$) whose radio emission is
dominated by the radio AGN and are considered as the RL counterpart of
the sample. With the sole exception of 625, they are associated
with red massive ETGs belonging to the LEG spectroscopic class,
similar to the vast majority of the B05 sample.

\subsection{Radio data}

\begin{figure}
\centerline{
\includegraphics[scale=0.15,angle=-90]{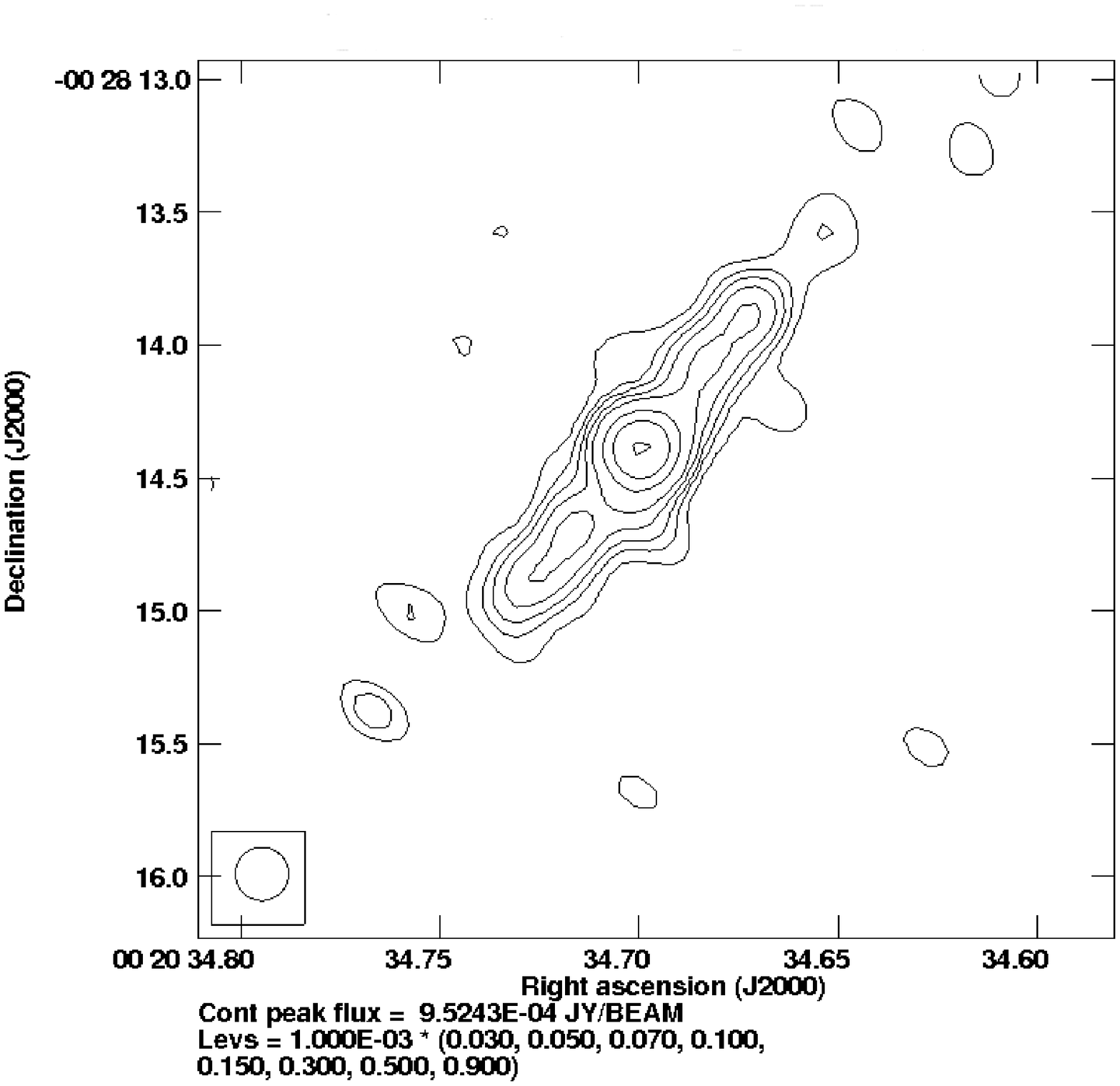}
\includegraphics[scale=0.15,angle=-90]{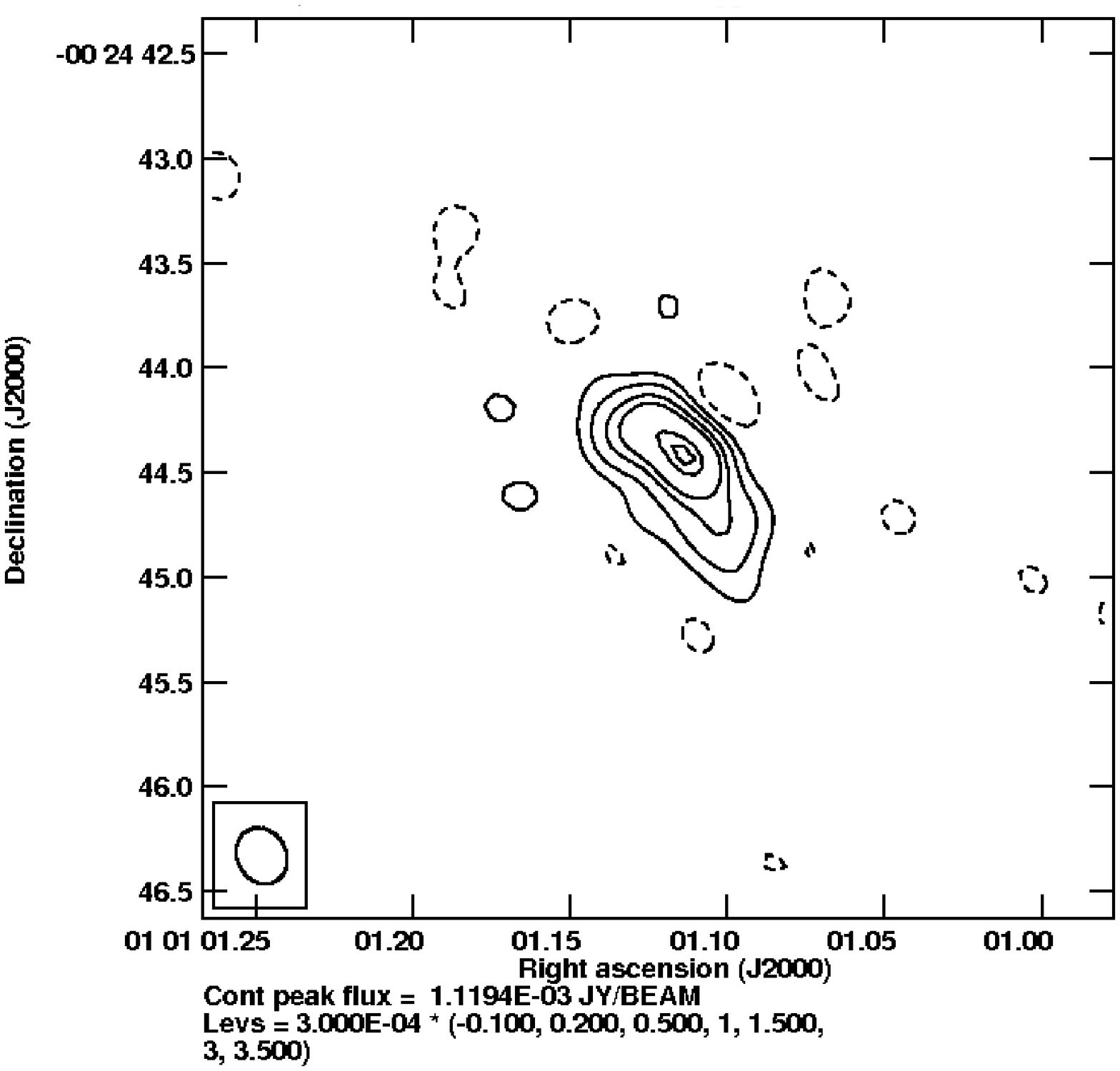}}
\centerline{
\includegraphics[scale=0.15,angle=-0.0]{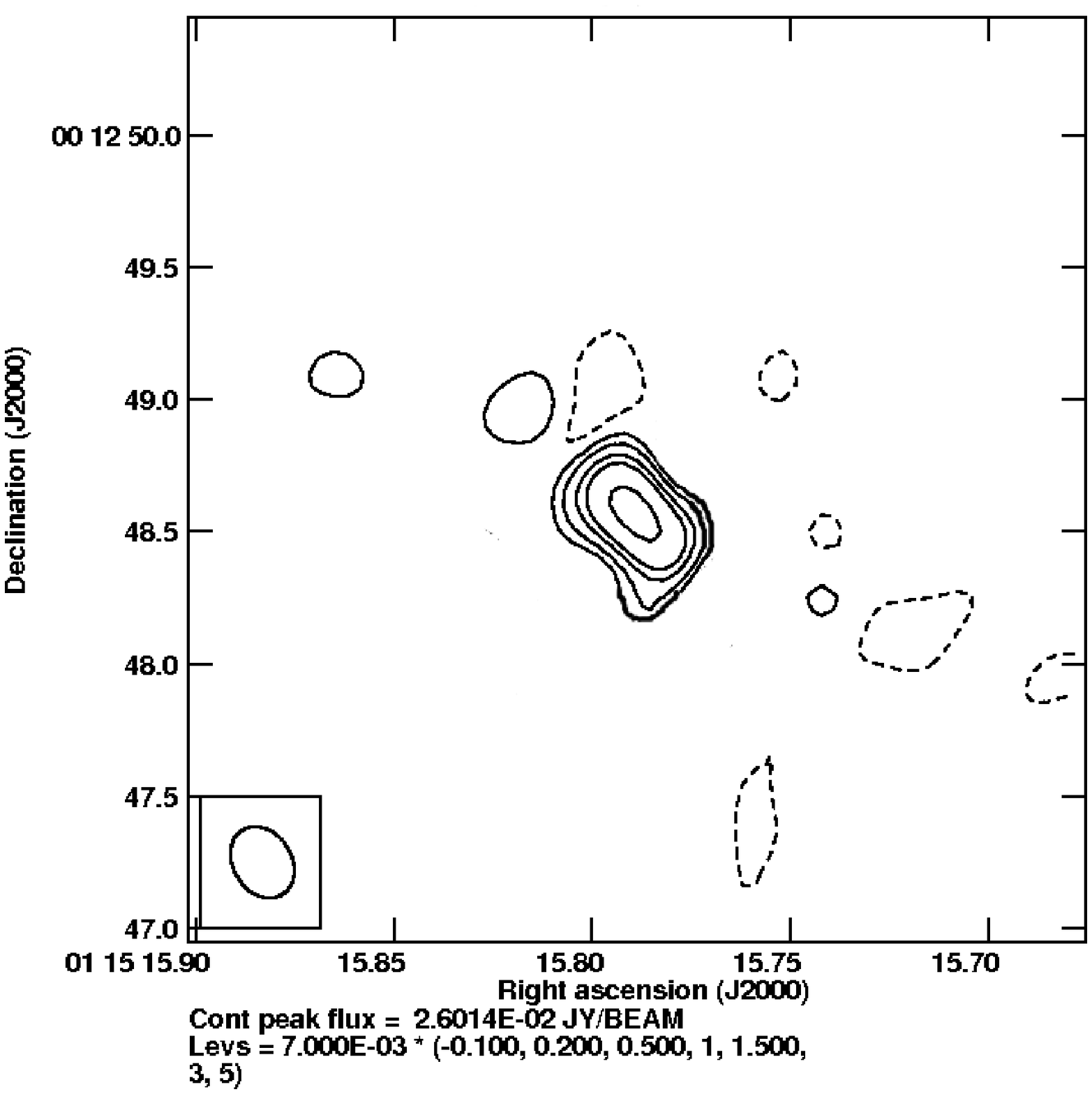}}
\caption{{\small{The JVLA maps at 7.5 GHz of the three FR~0s, (from left top 547, 590, and
      605) which show extended structure (resolution of $\sim$0\farcs2) . See
      \citet{baldi15a} for the expanded radio maps.}}}
\label{maps} 
\end{figure}

Let us focus on the new new high-resolution JVLA observations at 1.4,
4.5, and 7.5 GHz of the eight RL objects. One object, 625, show a
FRI/FRII radio morphology extended over $\sim$40kpc and has a HEG
optical spectrum. We exclude it from the final sample, consisting of 7
objects (see the luminosities and properties in
Table~\ref{table3}). The new JVLA maps of these objects revealed that
they show a compact morphology, unresolved or only slightly resolved
(core-jet) down to a resolution of $\sim$0\farcs2 (Fig.~\ref{maps}). The total
morphology show structure on a scale of 1-3 kpc. The sources which
show an interesting morphology are: 547 (a twin-jet source), and 590
(with an elongated radio morphology on a scale of 0\farcs8, possibly
due to a bent two-sided jet structure).

\begin{table*}
\caption{Spectroscopic and photometric properties of the FR~0s}
\begin{tabular}{l|c|c| c|c|c|c|c|c|}
\hline
  ID  & z & opt. class  &   Log M$_{*}$  & Log M$_{BH}$  & Log L$_{\rm [O~III]}$ &  Log L$_{FIRST}$  & Log L$_{NVSS}$  & Log L$_{\rm core}$ \\
\hline
  519 & 0.076 & LEG  & 11.30  &    8.67 &    40.28   &   39.50  &   39.56   &     39.31   \\
  524 & 0.093 & LEG  & 11.07  &    8.32 &    39.96   &   39.89   &   39.87   &     $<$40.16  \\
  535 & 0.076 & LEG  & 11.48  &    8.76 &    40.26   &   38.96   &   39.14   &     39.29     \\
  537 & 0.061 & LEG  & 11.09  &    7.72 &    39.52   &   38.90   &   38.90   &     39.29      \\
  547 & 0.072 & LEG  & 11.18  &    7.64 &    39.97   &   39.22   &   39.27   &     38.93        \\
  590 & 0.097 & LEG    & 11.42  &    8.43 &    40.39   &   39.22  &   39.37   &     39.64       \\
  605 & 0.045 & LEG  & 11.15  &    8.57 &    39.57   &   39.47   &   39.44   &     40.10       \\
\hline
\end{tabular}
\label{table3}

\medskip
\small{Column description: (1) name; (2) spectroscopic redshift; (3)
  optical spectroscopic classification; (4) galaxy stellar mass M$_{*}$ (M$_{\odot}$); (5) black
  hole mass M$_{BH}$ (M$_{\odot}$); (6) [O~III] luminosity (erg
  s$^{-1}$); (7) 1.4 GHz FIRST luminosity (erg s$^{-1}$); (8) 1.4
  GHz NVSS luminosity (erg s$^{-1}$) used as total radio luminosity
  L$_{\rm tot}$; (9) 7.5 GHz JVLA luminosity (erg s$^{-1}$) used as
  radio core luminosity L$_{\rm core}$.}
\end{table*}

We measure the spectral slope $\alpha$ ($F_{\nu} \propto
\nu^{\alpha}$), obtained between 1.4 and 4.5 GHz: they range from
-1.00 to -0.04, indicative of steep and flat spectra.  For most of the
objects, the spectra show a flattening at 7.5 GHz (Fig. 4 from
\citealt{baldi15a}, sign of an emerging radio core component.

One of the main purposes of our program is to isolate the radio core component
and to measure the core dominance of the sources of our sample. The
high-resolution radio maps at 7.5 GHz reveals an unresolved radio core
component for three sources (519, 537, and 547). For the remaining sources, we
can use the radio spectra to measure the core emission, at the frequency where
the spectra flatten. The radio core luminosities are in the range
$\sim$10$^{39-40}$ erg s$^{-1}$, similar to the cores of 3CR/FR~Is.

\begin{figure}
\centering
\includegraphics[scale=0.45,angle=0]{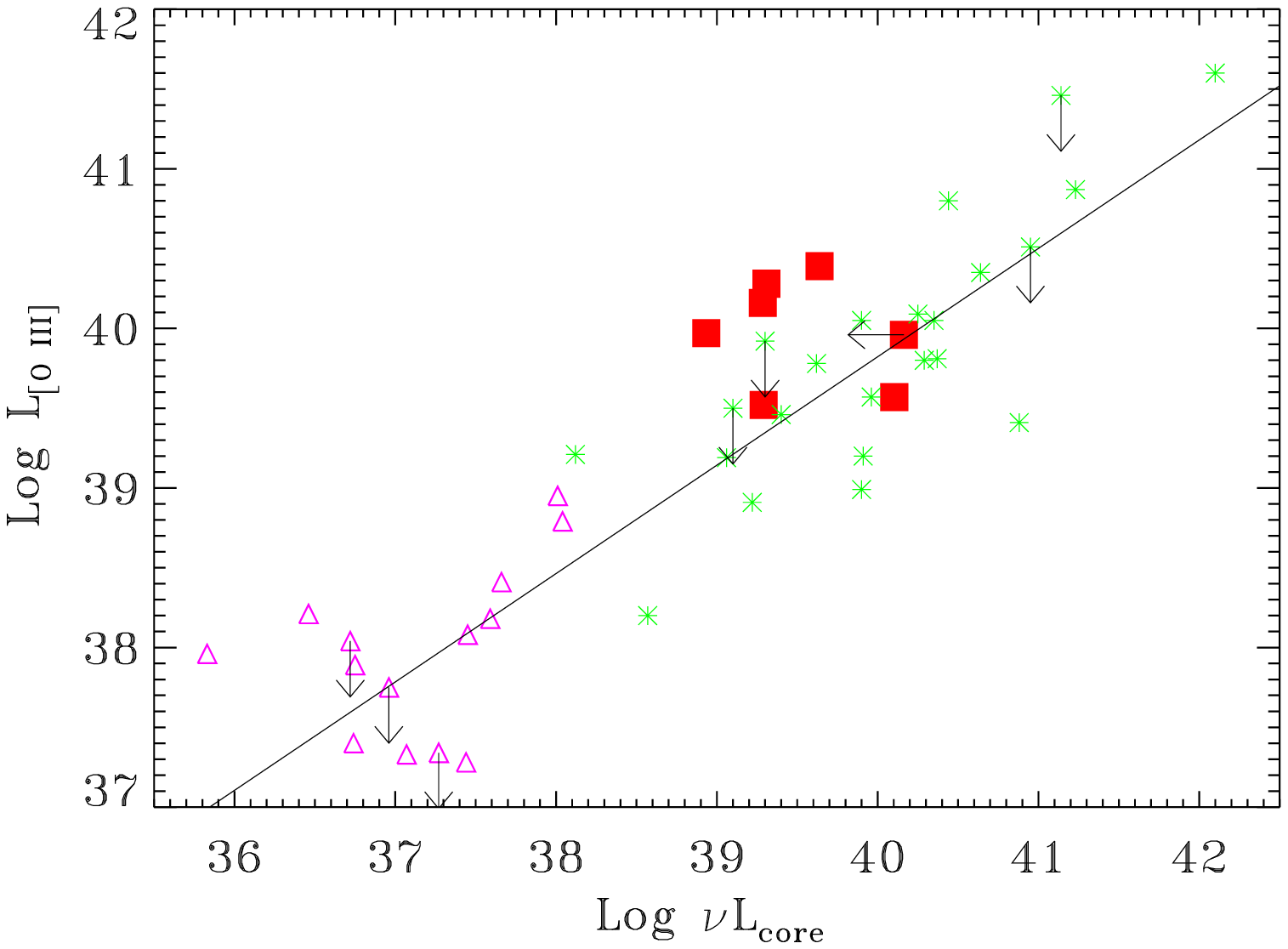}
\caption{{\small{Core radio power vs. [O~III] line luminosity (erg
      s$^{-1}$). The line indicates the best linear fit the for
      3CR/FR~I. The color symbols are like in Fig.~1}}}
\label{lcorelo3} 
\end{figure}

After measuring the radio core component thanks to the new
observations, we can include our sources in the $L_{core}$
vs. \loiii\ plane (Fig.~\ref{lcorelo3}), similarly to what was done
for CoreG in \citet{baldi09}. The seven sources lie on the $L_{core}$
vs. \loiii\ correlation found for the 3CR/FR~Is. This strongly
indicates their radio-loudness and their genuine lack of substantial
large-scale radio emission since no geometric effect is present
(Fig~\ref{lrlo3}).

The core dominance of our sources is defined as the ratio between the
source nuclear emission at 7.5 GHz and the total flux density, for
which we adopted the NVSS measurement. The core dominance ranges
between 0.05 and 0.86 (Fig. 7 from \citealt{baldi15a}) and is
consistent with the core dominance of CoreG and higher than 3CR/FR~Is
by a factor $\sim$30.

\section{Discussion}

The results discussed in the previous sections indicate that these
seven objects (together with the powerful source 625 discussed above)
are the genuine RL AGN in the pilot sample selected from B05. They are
located in red massive ($\sim$10$^{11}$ M$_{\odot}$) ETGs, have BH
masses $\gtrsim$10$^{8}$ M$_{\odot}$ \citep{chiaberge11} and are
spectroscopically classified as LEGs. All these (host and nuclear)
properties are shared with FR~I RGs (e.g.,
\citealt{chiaberge99,baldi08,baldi09,balmaverde06,baldi15b}). Furthermore,
their radio core and [O~III] luminosities lie in the range typical of
FR~Is. The JVLA observations show compact radio structures (with a
limit to their size of $\sim$ 0.5 kpc or, in a few cases, extending by
at most 1-3 kpc) and lead to an estimate of their (average) core
dominance a factor of $\sim$30 higher than FR~Is.  The only feature
distinguishing them from FR~Is is then the substantial lack of
extended radio emission. For this paucity we call them ’{\it FR~0}’ in
opposition to the jetted FR radio classes \citep{fanaroff74} and in
agreement with \citet{ghisellini11}.

\subsection{The FR~0 population}

The FR~0 classification then corresponds to a combination of radio and
spectro-photometric (both nuclear and of the host) properties. FR~0s
and FR~Is are indistinguishable in terms of nuclear and host
properties (color, BH mass, optical spectra). They show compact radio
structures on scale of some kpc. The high FR~0 core dominance appears
to be due to the genuine paucity of extended radio emission, rather
than to an enhanced radio core, since FR~0s and FR~Is show similar
ratios of radio-core to emission line luminosity.

The CoreG fulfill the requirements for a FR~0 classification: they
show kpc scale radio structures and are of high core dominance, they
are hosted in red giant ellipticals and are characterized by LEG line
ratios.  They are $\sim$100 times less luminous than the FR~0s studied
here. They smoothly extend the various nuclear multi-wavelengths
relations seen in FR~Is. In this sense, they represent the
low-luminosity end of the FR~0 population that therefore extends at
least down to a radio power of $\sim 10^{36}$ $\ergs$.

Since the vast majority of the B15 sample fulfills the FR~0 definition
(a deficit of radio emission and similar spectro-photometric
properties to FR~Is), this strongly indicates that the FR~0 population
is the dominant radio class in the local Universe. However, a detailed
radio study of a large portion of this sample is needed to put these
results on a firmer ground. This conclusion was also recently claimed
by \citet{sadler14}, who found that the bulk of the 20-GHz RG
population consists of compact radio sources lacking extended radio
emission, analogous to our FR~0s.

This new class of RGs is similar to the low-luminosity radio sources
hosted in ETGs, studied by \citet{slee94}, which contain parsec-scale
radio cores and do not produce extended radio emission. The presence
of a large population of low luminosity compact sources has been
recently unveiled by \citet{kunert10a}, similar to FR~0s but much
brighter. This new RG class is also consistent with a subclass of
low-luminosity Gigahertz-Peaked Spectrum (GPS) radio sources proposed
by \citet{tingay15}, which show jet-dominated compact morphologies
similar to FR~Is, but lacking extended radio emission.

What causes the radio deficit in FR~0s? In CoreG small scale jets and
plumes are the dominant radio morphology. This favors the idea that
their jets suffer deceleration and disruption before escaping the host
core radius, slowly burrowing their way into the external medium and
accounting for their small sizes. Furthermore, the similarity of the
host and nuclear properties of FR~Is and FR~0s constrain two scenarios
that can explain their extended radio difference.

In the first scenario, the central engines of FR~0s and FR~Is are
indistinguishable and the paucity (and small size) of the extended
radio emission is ascribed to an evolutionary effect. Since FR 0s
appear small and compact, they might be young and will possibly evolve
into more extended FR~I and FR~II. However, this implies that their
number density should be much lower than that of FR Is, but this is
the opposite of what is observed. This scenario can still hold if
FR~0s are intermittent sources.  Rapid intermittency, e.g., with a
timescale of $\sim$10$^{4-5}$ years would prevent FR~0s from becoming
well developed FR~Is. Several suggestions in this direction have been
proposed (e.g. \citealt{reynolds97})
based on various physical mechanisms, such as disk (or jet)
instabilities or discontinuous accretion. The drawback of this
scenario is that it does not explain why these mechanisms should be at
work only in FR~0s. Furthermore, this scenario is in contradiction
with the studies on FR~Is and CoreG \citep{allen06,balmaverde08},
which indicate that accretion is provided by the X-ray emitting hot
gas of the corona, which slowly cools, shrinks and supplies the
central BH. This implies a long-lasting continuous inward transfer of
gas related only to the host properties and then no differences would
be expected between FR~0s and FR~Is.

As a second scenario, we suggest that the differences between FR~0s
and FR~Is are driven by a different value of the jet bulk speed
$\Gamma$, with FR~0s having lower $\Gamma$ values with respect to
FR~Is. In this scheme, FR~0s and FR~Is share a common range of
accretion rate (as proved by their similar line emission luminosity
and similar optical line ratios) and of radio core power (which
represents the synchrotron emission from the base of the
jet). Therefore, the innermost regions of FR~0 and FR~Is are not
expected to differ significantly. The different radio behaviors should
arise on a larger scale. In the hypothesis that jets in FR~0s are
slower than FR~Is, they are more subject to instabilities and
entrainment \citep{perucho12,bodo13} and this causes their premature
disruption.  Indeed, the typical scale of the radio emission in FR~0s
is usually smaller than the core size of their host galaxies coronae,
a region characterized by a dense interstellar medium that easily may
obstruct the passage of the jet.  This hypothesis is supported, albeit
with a small number statistics, by the absence of one-sided kpc scale
morphologies among the FR~0s observed, the typical sign of
relativistic jet boosting.

The ultimate origin of this effect is apparently not related to any
directly observable quantity. We speculate that this could be due to a
different spin of their central BH. A broad and continuous
distribution of BH spin is a indeed natural consequence of galaxy
evolution via both BH mergers and gas accretion (e.g.,
\citealt{volonteri13}). By assuming that dependence between the BH
spin (and the BH mass) and jet bulk Lorentz factor $\Gamma$ exists
(e.g., \citealt{chai12,liu15}), this would result in the needed spread
of $\Gamma$ values. The FR~Is might represent the cases where the
efficiency in the mechanism of launching of a relativistic jet is
maximized. The FR~I radio morphology is produced only when the BH spin
is close to its maximum value, while smaller spin values could be
associated with FR~0s. Another benefit of this scenario is that, since
FR~Is represent the cases with extreme values of spin, the broad range
of BH spins would account for the more numerous population of FR~0s
than FR~Is. Circumstantial evidence in favor of this speculation comes
from the slight differences in large scale environment between FR~0s
and FR~Is. The latter class is generally ($\gtrsim$ 70\%) associated
with clusters or rich groups (e.g. \citealt{wing11}). Conversely,
although FR~0s avoid regions of low galaxies density and are located
in a richer environment than RQ AGN (B05), they are also found outside
clusters. This could alter slightly their evolution with respect to
FR~Is (i.e., from the point of view of merger rate and/or merger
type), leading to a difference in the BH spin distribution.

\section{Summary and Conclusions}

The most studied catalogues of RGs are severely biased against the
inclusion of objects with high core dominance, since a large
contribution from extended emission is needed to fulfill the stringent
flux requirements of low frequency, high flux threshold samples.
Conversely, recent studies on low-luminosity RGs (e.g.,
\citealt{baldi09,baldi10a,sadler14,baldi15a}) have been converging to
an opposite picture: the local RL AGN population is dominated by
compact high core-dominated weak RGs, the FR~0s, which show a lack of
prominent extended radio structures with respect to other FR
classes. Such a dominant population is still unexplored, casting
shadow on our present knowledge about the radio-AGN phenomena.

The FR~0 radio structures are not small because of a low-power engine
or because of geometrical effects, and their different radio
properties cannot be ascribed to differences in their hosts. Our
preferred scenario to account for their radio behavior is the slow jets,
associated with small jet Lorentz factors ($\Gamma \lesssim 2-3$),
probably due to small BH spins.  Furthermore, this conjecture matches
with the idea of recent studies on the radio-mode feedback which have
being gradually oriented to low-power/compact jets
(e.g. \citealt{shabala11,perucho14,cielo14}), because they appear to
be more efficient to deposit energy on the galaxy scale rather than
powerful radio galaxies which flow faster out from the galaxy.




\begin{thebibliography}{}

\end{thebibliography}


\begin{thebibliography}{}
{\footnotesize
\bibitem[{{Allen} {et~al.}(2006){Allen}, {Dunn}, {Fabian}, {Taylor}, \&
  {Reynolds}}]{allen06}
{Allen}, S.~W., {Dunn}, R.~J.~H., {Fabian}, A.~C., {Taylor}, G.~B., \&
  {Reynolds}, C.~S. 2006, \mnras, 372, 21

\bibitem[{{Baldi} \& {Capetti}(2008)}]{baldi08}
{Baldi}, R.~D. \& {Capetti}, A. 2008, \aaa, 489, 989

\bibitem[{{Baldi} \& {Capetti}(2009)}]{baldi09}
{Baldi}, R.~D. \& {Capetti}, A. 2009, \aaa, 508, 603

\bibitem[{{Baldi} \& {Capetti}(2010)}]{baldi10a}
{Baldi}, R.~D. \& {Capetti}, A. 2010, \aaa, 519, A48+

\bibitem[{{Baldi} {et~al.}(2010){Baldi}, {Chiaberge}, {Capetti}, {Sparks},
  {Macchetto}, {O'Dea}, {Axon}, {Baum}, \& {Quillen}}]{baldi10b}
{Baldi}, R.~D., {Chiaberge}, M., {Capetti}, A., {et~al.} 2010, \apj, 725, 2426

\bibitem[Baldi et al.(2015)]{baldi15a} Baldi, R.~D., Capetti, A., \& Giovannini, G.\ 2015, \aaa, 576, A38 

\bibitem[Baldi et al.(2015)]{baldi15b} Baldi, R.~D., Giroletti, M., Capetti, A., et al.\ 2015, \aaa, 574, A65 


\bibitem[{{Baldwin} {et~al.}(1981){Baldwin}, {Phillips}, \&
  {Terlevich}}]{baldwin81}
{Baldwin}, J.~A., {Phillips}, M.~M., \& {Terlevich}, R. 1981, \pasp, 93, 5

\bibitem[{{Balmaverde} \& {Capetti}(2006)}]{balmaverde06core}
{Balmaverde}, B. \& {Capetti}, A. 2006, \aaa, 447, 97

\bibitem[{{Balmaverde} {et~al.}(2006){Balmaverde}, {Capetti}, \&
  {Grandi}}]{balmaverde06}
{Balmaverde}, B., {Capetti}, A., \& {Grandi}, P. 2006, \aaa, 451, 35

\bibitem[{{Balmaverde} {et~al.}(2008){Balmaverde}, {Baldi}, \&
  {Capetti}}]{balmaverde08}
{Balmaverde}, B., {Baldi}, R.~D., \& {Capetti}, A. 2008, \aaa, 486, 119

\bibitem[{{Balogh} {et~al.}(1999){Balogh}, {Morris}, {Yee}, {Carlberg}, \&
  {Ellingson}}]{balogh99}
{Balogh}, M.~L., {Morris}, S.~L., {Yee}, H.~K.~C., {Carlberg}, R.~G., \&
  {Ellingson}, E. 1999, \apj, 527, 54

\bibitem[{{Best} {et~al.}(2005){Best}, {Kauffmann}, {Heckman}, \&
  {Ivezi{\'c}}}]{best05a}
{Best}, P.~N., {Kauffmann}, G., {Heckman}, T.~M., \& {Ivezi{\'c}}, {\v Z}.
  2005, \mnras, 362, 9

\bibitem[{{Bodo} {et~al.}(2013){Bodo}, {Mamatsashvili}, {Rossi}, \&
  {Mignone}}]{bodo13}
{Bodo}, G., {Mamatsashvili}, G., {Rossi}, P., \& {Mignone}, A. 2013, \mnras,
  434, 3030

\bibitem[{{Buttiglione} {et~al.}(2010){Buttiglione}, {Capetti}, {Celotti},
  {Axon}, {Chiaberge}, {Macchetto}, \& {Sparks}}]{buttiglione10}
{Buttiglione}, S., {Capetti}, A., {Celotti}, A., {et~al.} 2010, \aaa, 509, A6+

\bibitem[{{Chai} {et~al.}(2012){Chai}, {Cao}, \& {Gu}}]{chai12}
{Chai}, B., {Cao}, X., \& {Gu}, M. 2012, \apj, 759, 114

\bibitem[{{Chiaberge} {et~al.}(1999){Chiaberge}, {Capetti}, \&
  {Celotti}}]{chiaberge99}
{Chiaberge}, M., {Capetti}, A., \& {Celotti}, A. 1999, \aaa, 349, 77

\bibitem[{{Chiaberge} \& {Marconi}(2011)}]{chiaberge11}
{Chiaberge}, M. \& {Marconi}, A. 2011, \mnras, 416, 917

\bibitem[Cielo et al.(2014)]{cielo14} Cielo, S., 
Antonuccio-Delogu, V., Macci{\`o}, A.~V., Romeo, A.~D., 
\& Silk, J.\ 2014, \mnras, 439, 2903 


\bibitem[{{Fanaroff} \& {Riley}(1974)}]{fanaroff74}
{Fanaroff}, B.~L. \& {Riley}, J.~M. 1974, \mnras, 167, 31P

\bibitem[Ghisellini(2011)]{ghisellini11} Ghisellini, G.\ 2011, 
American Institute of Physics Conference Series, 1381, 180 


\bibitem[Kunert-Bajraszewska et al.(2010)]{kunert10a} 
Kunert-Bajraszewska, M., Gawro{\'n}ski, M.~P., Labiano, A., 
\& Siemiginowska, A.\ 2010, \mnras, 408, 2261 

\bibitem[Liu et al.(2015)]{liu15} Liu, Z., Yuan, W., Lu, Y., 
\& Zhou, X.\ 2015, \mnras, 447, 517 


\bibitem[{{Morganti} {et~al.}(1997){Morganti}, {Oosterloo}, {Reynolds},
  {Tadhunter}, \& {Migenes}}]{morganti97}
{Morganti}, R., {Oosterloo}, T.~A., {Reynolds}, J.~E., {Tadhunter}, C.~N., \&
  {Migenes}, V. 1997, \mnras, 284, 541

\bibitem[Perucho(2012)]{perucho12} Perucho, M.\ 2012, 
International Journal of Modern Physics Conference Series, 8, 241 

\bibitem[Perucho et al.(2014)]{perucho14} Perucho, M., 
Mart{\'{\i}}, J.~M., Laing, R.~A., 
\& Hardee, P.~E.\ 2014, \mnras, 441, 1488 

\bibitem[{{Reynolds}(1997)}]{reynolds97}
{Reynolds}, C.~S. 1997, \mnras, 286, 513


\bibitem[{{Sadler} {et~al.}(2014){Sadler}, {Ekers}, {Mahony}, {Mauch}, \&
  {Murphy}}]{sadler14}
{Sadler}, E.~M., {Ekers}, R.~D., {Mahony}, E.~K., {Mauch}, T., \& {Murphy}, T.
  2014, \mnras, 438, 796

\bibitem[Shabala et al.(2011)]{shabala11} Shabala, S.~S., 
Kaviraj, S., \& Silk, J.\ 2011, \mnras, 413, 2815 


\bibitem[{{Shen} {et~al.}(2003){Shen}, {Mo}, {White}, {Blanton}, {Kauffmann},
  {Voges}, {Brinkmann}, \& {Csabai}}]{shen03}
{Shen}, S., {Mo}, H.~J., {White}, S.~D.~M., {et~al.} 2003, \mnras, 343, 978

\bibitem[{{Slee} {et~al.}(1994){Slee}, {Sadler}, {Reynolds}, \&
  {Ekers}}]{slee94}
{Slee}, O.~B., {Sadler}, E.~M., {Reynolds}, J.~E., \& {Ekers}, R.~D. 1994,
  \mnras, 269, 928


\bibitem[Tingay \& Edwards(2015)]{tingay15} Tingay, S.~J., \& Edwards, P.~G.\ 2015, \mnras, 448, 252 

\bibitem[{{Tremaine} {et~al.}(2002){Tremaine}, {Gebhardt}, {Bender}, {Bower},
  {Dressler}, {Faber}, {Filippenko}, {Green}, {Grillmair}, {Ho}, {Kormendy},
  {Lauer}, {Magorrian}, {Pinkney}, \& {Richstone}}]{tremaine02}
{Tremaine}, S., {Gebhardt}, K., {Bender}, R., {et~al.} 2002, \apj, 574, 740



\bibitem[Volonteri et al.(2013)]{volonteri13} Volonteri, M., 
Sikora, M., Lasota, J.-P., \& Merloni, A.\ 2013, \apj, 775, 94 

\bibitem[Wing 
\& Blanton(2011)]{wing11} Wing, J.~D., \& Blanton, E.~L.\ 2011, \aj, 141, 88 


}
\end{thebibliography}

\end{document}